\documentclass[twocolumn,aps,a4paper,showpacs,superscriptaddress,amsmath,amssymb]{revtex4}

\usepackage{graphicx}
\usepackage{dcolumn}

\newcommand{\compA}{$\mathrm{Ga}_{1-x}\mathrm{Mn}_{x}\mathrm{As}$}
\newcommand{\compB}{(Ga,Mn)As}

\newcommand{\mni}{$\mathrm{Mn}_{I}$}
\newcommand{\asga}{$\mathrm{As}_{Ga}$}
\newcommand{\tc}{$T_{C}$}
\newcommand{\Ta}{$T_{a}$}
\newcommand{\ta}{$t_{a}$}
\newcommand{\tg}{$T_{g}$}
\newcommand{\kc}{$K_{c}$}
\newcommand{\ku}{$K_{u}$}
\begin{document}
%
    \title{Annealing-induced changes of the magnetic anisotropy of
    \compB\ epilayers}

    \author{V. Stanciu}
        \email{victor.stanciu@angstrom.uu.se}
        \affiliation{Department of Engineering Sciences, Uppsala University, Box 534, SE-751 21 Uppsala, Sweden}
    \author{P. Svedlindh}
        \affiliation{Department of Engineering Sciences, Uppsala University, Box 534, SE-751 21 Uppsala, Sweden}


\date{\today}
    \begin{abstract}
The dependence of the magnetic anisotropy of As-capped \compB\
epilayers on the annealing parameters - temperature and time - has
been investigated. A uniaxial magnetic anisotropy is evidenced,
whose orientation with respect to the crystallographic axes
changes upon annealing from $[\bar{1}10]$ for the as-grown samples
to $[110]$ for the annealed samples. Both cubic an uniaxial
anisotropies are tightly linked to the concentration of charge
carriers, the magnitude of which is controlled by the annealing
process.
    \end{abstract}
    \maketitle
%
Considerable experimental and theoretical efforts have recently
been devoted to the study of III-V diluted magnetic semiconductors
(DMS) due to their potential implementation as spintronic devices
\cite{Wolf:Science-294-1488-2001}. The high ferromagnetic
transition temperature of \compB\ has made this compound one of
the most investigated III-V DMS, being regarded as the prototype
of this new class of materials \cite{MacDonald:NM-4-195-2005}. It
is well established nowadays that the ferromagnetic interaction
between the magnetic ions is mediated by the charge carriers,
holes in this case, and that point defects such as As antisites
(\asga) and Mn interstitials (\mni) play a crucial role in
determining the magnetic properties of \compB\
\cite{Yu:PRB-65-201303-2002,Ku:APL-82-2302-2003,Edmonds:PRL-92-037201-2004}.
Optimal annealing experiments have proved that the highly mobile
\mni\ defects \cite{Yu:PRB-65-201303-2002} out-diffuse to and are
passivated at the free surface
\cite{Edmonds:PRL-92-037201-2004,Stanciu:AnnealingPaper} and that
in all successful annealings the \mni\ concentration is higher
than the concentration of \asga. The removal of \mni\ from the
bulk of \compB\ leads to an increase of the ferromagnetic
transition temperature (\tc), carrier concentration and average
manganese magnetic moment.
\begin{figure}
        \centering
        \includegraphics[width=0.48\textwidth]{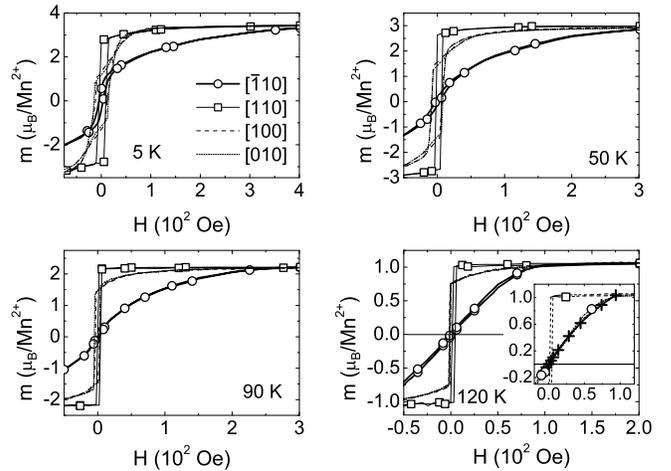}
        \caption{\label{fig:fig1} Field dependence of the
        magnetization recorded at different temperatures for a 1000~\AA~thick
        sample annealed at 240~$^\circ\mathrm{C}$ for 10h. The crosses in the inset
        of the `120K' pane represent a fit to the hard axis magnetization
        as discussed in the text.}
\end{figure}
However, one particular aspect of the magnetic properties is not
fully understood and this concerns the magnetic anisotropy.
Experimentally it is found that a strong uniaxial (UA)
contribution to the overall magnetic anisotropy appears along the
$[110]$ direction
\cite{Welp:PRL-90-167206-2003,Welp:APL-85-260-2004,Sawicki:PRB-71-121302-2005}.
Although the \compB\ structure is tetragonally distorted due to
its lattice mismatch with the GaAs substrate, structural
characterization commonly shows that the epilayers are coherently
strained to the substrate \cite{Schott:APL-79-1807-2001} and this
holds even for very large thicknesses of the order of a few $\mu
\mathrm{m}$ \cite{Welp:APL-85-260-2004}. Theoretical
investigations explain fairly well the influence of the biaxial
strain and hole concentration on the magnetic anisotropy
\cite{Dietl:PRB-63-195205-2001,Abolfath:PRB-63-054418-2001}.
However, based on symmetry arguments the tetragonal distortion
cannot account for the appearance of the uniaxial anisotropy.
Anisotropy measurements on as-grown samples of different
thicknesses \cite{Welp:APL-85-260-2004} and on annealed-and-etched
samples \cite{Sawicki:PRB-71-121302-2005} have shown that the UA
contribution does not have a surface or interface origin. Instead,
it has been suggested by Sawicki~\emph{et al.}
\cite{Sawicki:PRB-71-121302-2005} that the UA originates from a
small trigonal distortion $\epsilon_{xy}\neq0$; using the $p-d$
Zener model of the ferromagnetism they estimate that
$\epsilon_{xy}=0.05$\%~is enough to explain the experimentally
observed UA. The microscopic origin of the trigonal distortion
could be surface As-dimerization, resulting in surface
reconstruction and an inequivalence between the $[110]$ and
$[\bar{1}10]$ directions \cite{Welp:PRL-90-167206-2003}.

The \compA\ ($x\sim0.06$) samples were grown on (001) GaAs
substrates by low-temperature molecular beam epitaxy at a growth
temperature (\tg) of about 230~$^\circ$C. Further details on the
sample preparation are given elsewhere \cite{Adell:APL}. For an
efficient annealing, a rather thick capping layer of amorphous As
was deposited on top of \compB. Pieces of the as-grown samples
were annealed in air at different annealing temperatures (\Ta) and
for different annealing times (\ta). We have shown that the
dependence of \tc\ on the annealing time has a peculiar variation
due to the presence of the amorphous cap
\cite{Stanciu:AnnealingPaper}. Depending on the `position' of \Ta\
with respect to \tg, \tc-peaks appear at certain values of \ta\
($t_a^{peak}$). In this study we focus only on \Ta\ either lower
(215~$^\circ\mathrm{C}$) or higher (240~$^\circ\mathrm{C}$) than
\tg. The annealing times are chosen relative to the $t_a^{peak}$
position in order to cover \ta-regions below and above
$t_a^{peak}$ \cite{Stanciu:AnnealingPaper}.

The magnetic anisotropy of the samples was assessed from
field-dependent measurements of the magnetization recorded in a
QuantumDesign MPMS-XL SQUID magnetometer. For a detailed
investigation of the magnetic anisotropy, hysteresis loops were
measured along four different crystallographic directions, namely
$[100]$, $[110]$, $[010]$ and $[\bar{1}10]$, and at different
temperatures, both for the as-grown and annealed samples. An
example of such hysteresis loops is shown in Fig.~\ref{fig:fig1}
for a 1000~\AA\ thick sample annealed for 10h at
240~$^\circ\mathrm{C}$. The \tc\ was derived from the temperature
dependence of the magnetization, $M(T)$, as the onset of the FM
order; \tc\ values for the different samples are summarized in
Table~\ref{tab:table1}.
\begin{table}
\caption{\label{tab:table1} \tc\ values for the 1000~\AA~and
300~\AA\ thick samples annealed at 215~$^\circ\mathrm{C}$ and
240~$^\circ\mathrm{C}$ for different \ta. The as-grown samples are
denoted by `a'.}
\begin{ruledtabular}
\begin{tabular}{c | c c c c c c c c c |}
 &&&&& $t=1000$~\AA &&&&\\
\hline
\Ta\ & a & $|$ && 215~$^\circ\mathrm{C}$ && $|$ && 240~$^\circ\mathrm{C}$ &\\
\hline
\ta\ && $|$ & 3h & 7h & 10h & $|$ & 1h & 4h & 10h\\
\hline
\tc\ (K) & 76 & $|$ & 127 & 136 & 132 & $|$ & 127 & 145 & 130 \\
\hline \hline
 &&&&& $t=300$\AA &&&&\\
\hline
\Ta \ & a & $|$ && 215~$^\circ \mathrm {C}$ && $|$ && 240~$^\circ \mathrm {C}$ &\\
\hline
\ta\  && $|$ & 1h & 5h & 9h & $|$ & 3h & 4h & 10h\\
\hline
\tc\ (K) & 82 & $|$ & 127 & 136 & 131 & $|$ & 134 & 122 & 126 \\
\end{tabular}
\end{ruledtabular}
\end{table}
The free energy density, considering a cubic, an uniaxial and a
Zeeman term, can be written as\ \ $e = K_c\sin^2(\theta)\cos
^2(\theta) + K_u \sin^2(\theta-\pi/4) - \mu_0 M_s H\cos (\theta -
\alpha_H)$, where $\theta$ and $\alpha_H$ give the orientation of
the saturation magnetization ($M_{s}$) and applied field with
respect to the $[100]$ direction, arbitrarily chosen as a
reference axis. In our measurements, $\alpha_H$ therefore assumes
the following values $0$, $\pi/4$, $\pi/2$ and $3\pi/4$.
Minimization of the free energy with respect to the angle
$\theta$, and for particular values of $\alpha_H$, gives relations
between the applied field and the magnetization as in
Eq.~\ref{eg:fit_hard_axis} ($\alpha_H = 3\pi/4$).
\begin{equation}
    \label{eg:fit_hard_axis}
    \frac{{2K_c }}{{\mu _0 M_s }}\left[ {2\left( {\frac{M}{{M_s }}}
    \right)^3  - \frac{M}{{M_s }}} \right] + \frac{{2K_u }}{{\mu _0
    M_s }}\frac{M}{{M_s }} = H
\end{equation}
The anisotropy constants, \ku\ and \kc, are found from fits to the
measured hard-axis magnetizations, with the hard-axis
corresponding to the $[110]$ and $[\bar{1}10]$ directions for the
as-grown and annealed samples, respectively. An example of such a
fit is shown in the inset of the `120 K' pane of
Fig.~\ref{fig:fig1} (the fit is marked by crosses). One should
stress that errors stemming from determination of the sample
volume (thickness and area) directly influence the value of the
saturation magnetization and as a result rather large relative
errors of $5\div10 \%$ are introduced when calculating the
anisotropy constants.
\begin{figure}
        \centering
        \includegraphics[width=0.48\textwidth]{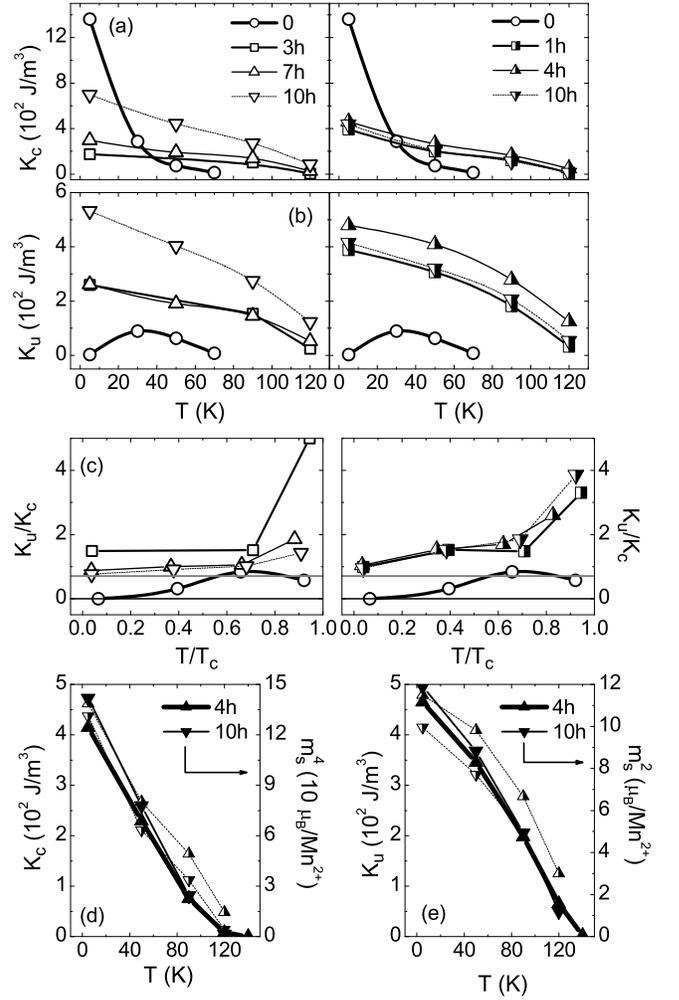}
        \caption{\label{fig:fig2} (a) and (b) \kc\ and \ku\ vs. temperature
        for a 1000 \AA-thick sample annealed at $215\ ^\circ\mathrm{C}$ (empty
        symbols, left hand panes) and $240\ ^\circ\mathrm{C}$ (half-filled symbols, right hand panes).
        (c) The ratio $K_u/K_c$ as a function of the reduce temperature $T/T_c$
        for $T_a=215\ ^\circ\mathrm{C}$ (left hand pane) and $T_a=240\ ^\circ\mathrm{C}$ (right hand pane).
        (d) $K_c$ (dashed lines) and $m_s^{4}$ (solid lines) vs. temperature ($T_a=240\ ^\circ\mathrm{C}$).
        (e) $K_u$ (dashed lines) and $m_s^{2}$ (solid lines) vs. temperature ($T_a=240\ ^\circ\mathrm{C}$).}
\end{figure}

\begin{figure}
        \centering
        \includegraphics[width=0.48\textwidth]{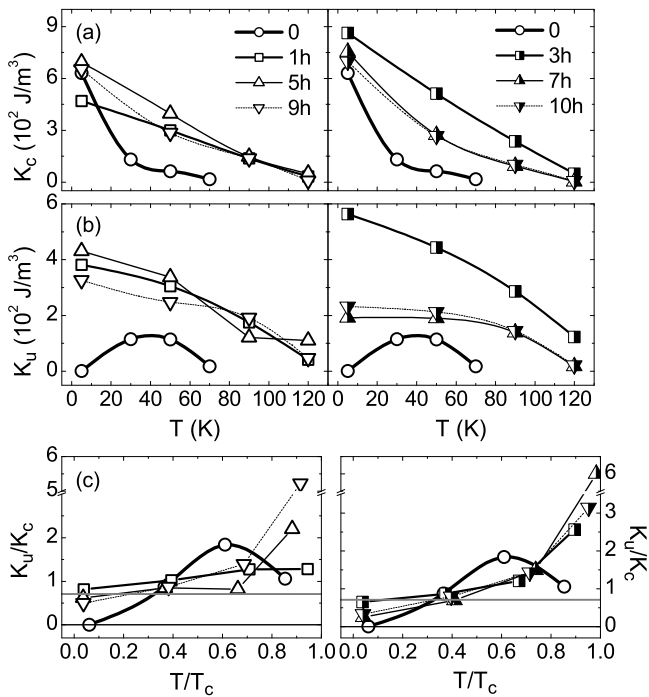}
        \caption{\label{fig:fig3} (a) and (b) \kc\ and \ku\ vs. temperature
        for a 300-\AA-thick sample annealed at $215\ ^\circ\mathrm{C}$ (empty
        symbols, left hand panes) and $240\ ^\circ\mathrm{C}$ (half-filled symbols, right hand panes).
        (c) The ratio $K_u/K_c$ as a function of the reduced temperature $T/T_c$
        for $T_a=215\ ^\circ\mathrm{C}$ (left hand pane) and $T_a=240\ ^\circ\mathrm{C}$ (right hand pane).}
\end{figure}

Figures~\ref{fig:fig2} and \ref{fig:fig3} show the temperature
dependence of the anisotropy constants for the 1000 \AA\ and 300
\AA\ thick films, respectively. The relative contribution of the
two anisotropies to the overall anisotropy is given by the ratio
$K_u/K_c$. This ratio is plotted in Figs.~\ref{fig:fig2}~(c) and
\ref{fig:fig3}~(c) for the two annealing temperatures. If $K_u/K_c
\geqslant 1$, the EA is pinned along $[110]$; if $0 \leqslant
K_u/K_c < 1$, the EA assumes intermediate directions between
$[110]$ and $[100]$; if $1/\sqrt{2} \leqslant K_u/K_c < 1$, the EA
is closer to $[110]$ and vice versa (the grey line in the figures
corresponds to $K_u/K_c = 1/\sqrt{2}$). In the limit case of a
cubic anisotropy, i.e. EA along $[100]$, $K_u$ tends to zero. All
these situations can be easily seen in the hysteresis loops. For
instance, in Fig.~\ref{fig:fig1}, at 5 K, $K_u/K_c = 0.9$, thus
the EA is very close to $[110]$, while at elevated temperatures,
as $K_u/K_c$ exceeds 1, square hysteresis loops are obtained with
$M_r^{[110]}\simeq M_s$, where $M_r$ is the remanence
magnetization. In Fig.~\ref{fig:fig2}~(d) and (e), one may notice
that \kc\ and \ku\ exhibit a temperature dependence in agreement
with that of $m_s^4$ and $m_s^2$, respectively, as expected for
the cubic and uniaxial anisotropy terms.

For the as-grown samples we find that at low temperatures the
uniaxial contribution is negligible as compared to cubic one; the
ratio $K_u/K_c$ approaches zero in this case, and thus an
equivalence between the \mbox{$\langle$100$\rangle$} and
\mbox{$\langle$110$\rangle$} axes is found in the hysteresis
loops. As the temperature increases, $K_c$ has a rapid fall-off
and an uniaxial contribution appears with the EA oriented
(roughly) along $[\bar{1}10]$. It should be noted, using the free
energy expression for the magnetization given above, that the
extracted value of \ku\ is negative for the as-grown samples.
However, in Figs.~\ref{fig:fig2}~(b) and \ref{fig:fig3}~(b) the
magnitude of \ku\ is plotted instead. Furthermore, the temperature
dependence of \ku\ is similar for both the $1000$ \AA~and $300$
\AA~thick samples, whereas \kc\ is larger for the $1000$
\AA~sample than for the $300$ \AA\ sample.

A 90-degree rotation of the UA takes place upon annealing. This
rotation occurs even for very short \ta\ signaling a strong
correlation between the UA direction and the hole concentration as
pointed out recently by Sawicki~\emph{et al.}
\cite{Sawicki:PRB-71-121302-2005}. For \Ta\ close to \tg\ short
annealing times are needed to readily deplete the bulk of \compB\
from \mni, and thus a large increase of the hole concentration is
expected \cite{Stanciu:AnnealingPaper}. Another manifestation of
the close connection between magnetic anisotropy and carrier
concentration is the variation of the anisotropy constants with
\ta\ since both \kc\ and \ku\ exhibit maxima at $t_a =
t_a^{peak}$; a maximum in hole concentration at $t_a = t_a^{peak}$
was inferred from the annealing experiments reported in
Ref.~\cite{Stanciu:AnnealingPaper}.

In summary, we have shown that both cubic and uniaxial
anisotropies appearing in \compB\ are influenced by changes in the
hole concentration, which in turn is controlled by the different
annealing parameters.

The Swedish Foundation for Strategic Research (SSF) and the Swedish
Research Council (VR) are acknowledged for financial support.
\bibliography{Anis_bibD5}
\end{document}